\documentclass[12pt]{spieman}  
\usepackage{amsmath,amsfonts,amssymb}
\usepackage{graphicx}
\usepackage{setspace}
\usepackage{tocloft}
\usepackage{lineno}
\usepackage{xcolor,soul}

\newcommand{\cmthree}{cm$^{-3}$}
\newcommand{\cmtwo}{cm$^{-2}$}
\newcommand{\adeg}{$^{\circ}$}
\newcommand{\arcmin}{$^{\prime}$}
\newcommand{\arcsec}{$^{\prime \prime}$}
\newcommand{\msun}{M$_{\odot}$}
\newcommand{\um}{$\mu$m}

\title{PRIMAGAL: a PRIMAger Galactic Plane Far-IR polarization survey to quantify the magnetic fields' role in the formation and evolution of large star-forming filaments.}

\author[a,*]{S.~Molinari}
\author[a]{J.~D.~Soler}
\author[a]{V.-M.~Pelkonen}
\author[a]{A.~Nucara}
\author[a]{E.~Schisano}
\author[a]{A.~Traficante}
\author[a]{C.~Mininni}
\author[a]{M.~Benedettini}
\author[a]{A.~Coletta}
\author[a]{D.~Elia}
\author[a]{S.~Pezzuto}
\affil[a]{Istituto Nazionale di Astrofisica (INAF)-Istituto di Astrofisica e Planetologia Spaziale, Via Fosso del Cavaliere 100, I-00133 Roma, Italy}

\cftpagenumbersoff{figure}
\cftpagenumbersoff{table} 
\begin{document} 
\maketitle

\begin{abstract}
The PRIMAger instrument on board the proposed PRIMA satellite will offer the unprecedented capability to obtain hundreds of square-degree maps in polarised emission at sub-arcminute resolution in four Far-IR bands. This will open a unique window to study magnetic fields in our Galaxy. PRIMAGAL, a proposed survey of polarized dust emission in the Milky Way Galactic Plane will determine the strength and orientation of magnetic fields towards several thousands of filamentary clouds in a wide range of linear masses, column densities, evolution, star-formation rates and efficiencies, and Galactic environment. This survey will address for the first time in a statistically significant fashion the role that magnetic fields play in shaping the formation, evolution and fragmentation of dense ISM filaments down to a minimum scale of 0.4 pc up to 8 kpc distance from the Sun. A 4-band polarization survey of the Galactic Plane with $\lvert$b$\rvert\leq 1$\adeg\ (a total of 720 sq. deg.) can be executed by PRIMAger in $\sim$1200 hours including all mapping and instrument overhead. 
\end{abstract}

\keywords{star formation, molecular clouds, filaments, magnetic field, polarization, shocks}

{\noindent \footnotesize\textbf{*}Sergio Molinari,  \linkable{sergio.molinari@inaf.it} }

{\noindent \footnotesize To appear in \textit{Journal of Astronomical Telescopes, Instruments and Systems} (JATIS), Vol. 11, Issue 3 (May 2025) }

\begin{spacing}{1}   

\section{Introduction}
\label{intro}  


Magnetic fields are important agents in shaping the way the diffuse Interstellar Medium (ISM) condense in large-scale filamentary clouds that then fragment to form the parsec-scale dense clumps that host nascent protoclusters. Early evidence from Zeeman-splitting line observations \cite{Crutcher+2010} suggested that while in relatively diffuse ISM the strength of the magnetic field is constant with gas density, it showed a rise with increasing density close to the criticality regime in terms of mass/flux ratio, after a threshold value roughly located at $10^2\leq n_H \leq 10^3$ \cmthree. This possibly hinted at different roles played by B in different stages of the formation process of dense structures. Curiously, this density transition regime is similar to the average density values for the large filamentary structures that permeate the Galactic Plane as revealed \cite{Schisano+2014, Schisano+2020} by the Herschel Hi-GAL survey \cite{Molinari2010b}. 

The importance of the magnetic field for the filamentary cloud formation was later revealed by the {\sl Planck} 353\,GHz polarization maps, which showed a fundamental statistical relationship between cloud structure and magnetic field orientation in 10 nearby ($d\,<500\,$pc) molecular clouds \cite{Soler-Planck2016}. In the surroundings of dense filamentary clouds the magnetic field lines tend to be oriented perpendicular to the backbone axis of clouds, and parallel to the faint filamentary pattern observed in their low column density environment. The orientation changes to parallel the closer one gets to the denser cloud regions. 
Simulations indicate that clouds only show this change in relative orientation when their magnetic fields are dynamically important; that is, when the magnetic energy density is equal to or larger than the turbulent kinetic energy density \cite{Soler+2013}. 

The spatial resolution of the {\sl Planck} maps in polarization ($\sim$5\arcmin\ in the Galactic Plane), however, was not sufficient to map magnetic fields across the transition region where magnetic field may become subdominant with respect to gravity. To understand how to role of B changes with clouds density and morphology it is crucial to probe polarization down at subarcminute level. 

An important leap in this direction was offered by facilities that allowed such resolution either because of large telescope apertures, such as Pol-2 at the JCMT \cite{Friberg+2016}, or because of shorter wavelengths of observations, like HAWC+ on board SOFIA Observatory \cite{Harper+2018}. Polarization observations with these two instruments showed \cite{Pillai+2020, Liu+2018} that the magnetic field may change back to aligning parallel to dense filaments as field lines are altered by gravitational collapse. However, the sensitivity and survey speed of HAWC+ and POL-2 make it impossible to survey large areas encompassing entire molecular cloud complexes. 

The PRIMAger instrument (Ciesla et al., this Volume) on board of the PRIMA observatory (Glenn et al., this volume) offers the unique and unprecedented combination of sensitivity, spatial resolution and mapping speed to effectively trace the strength and morphology of magnetic fields from the scale of entire large filamentary complexes down to the dense core scale.

\section{Mapping magnetic fields towards the Galactic Plane: uniqueness of PRIMAger}

 While the PRIMAger capabilities can be used to map in great detail the magnetic field in a few relatively nearby star-forming complexes (Pattle et al., Louvet et al., this Volume), we here propose to deploy PRIMAger's power to map several thousands of filamentary structures in the Milky Way Galactic Plane to explore in statistically significant fashion the role of magnetic fields in the assembly process of dense filamentary structures and their subsequent evolution and fragmentation in the dense parsec-scale clumps that are the sites for the formation of new stellar clusters. The Hi-GAL survey has unveiled the dense web of dusty filamentary structures in which the ISM gets organised along the path of forming stars \cite{Schisano+2020}. 
 A systematic survey like the one we propose will target (see below) relatively low column density filaments associated mostly with an atomic gas component and that are likely signposts for the first stages of ISM assembly into dense molecular clouds, as well as denser and actively star-forming filaments. 
  
The polarization mapping capabilities offered by the PRIMAger instrument fill a critical gap in sensitivity and spatial resolution between Planck on one side, that delivered all-sky mapping but at a too coarse ($\sim5$\arcmin) spatial resolution, and ALMA on the other with excellent sub-arcsecond resolution but very limited mapping power. With the suppression of the SOFIA observatory, PRIMAger is the only available to obtain a full Galaxy-wide characterization of the magnetic field properties towards and around hundreds-of-parsecs clouds down to the sub-parsec scales of dense clumps. With a beam FWHM between 9\arcsec\ and 24\arcsec\ in its 90-230\um\ spectral range (Ciesla et al., this Volume), PRIMAger will be able to resolve linear scales down to $\sim0.35$ pc up to $\sim$8 kpc heliocentric distance (i.e., the distance to the Galactic Center).

The full-sky polarization map from Planck reveals a Galactic Plane that within a few degrees from the midplane is dominated by an essentially uniform horizontal magnetic field that at the resolution of $\sim$5\arcmin\ shows very limited variations irrespective of the underlying ISM density structure \cite{Bernard+2015}. This is presumably due to the beam-filling integrated column of the diffuse ISM along the line of sight, which dominates the contribution of polarized emission from the denser and more compact filamentary structures that have a much lower filling factor in such a large beam.

To illustrate the unique potential that PRIMAger offers to make dramatic progress in this field, we post-processed in polarization full-Galaxy numerical simulations to produce synthetic maps of Stokes Q and U parameters that would be produced assuming PRIMAger spatial resolution. We started from an improved version of the MHD simulations that uses a barred potential to mimic the conditions of a Milky Way galaxy \cite{Tress+2024}. These simulations have a mass resolution of 20\,\msun\ within a 5\,kpc area of the Galactic center, and in a 1 kpc wide ring located at the Solar circle. Elsewhere, the mass resolution is 200\,\msun, which for dense gas is largely sub-parsec in spatial resolution, and thus is sufficient to fully capture the structure of dense clouds.

The post-processing of the MHD simulation is done with the POLARIS radiative transfer code \cite{Reissl+2016}. The dust composition is 62.5\% silicate and 37.5\% graphite in mass, with a size distribution of typical Milky Way dust: minimum size 5\,nm, maximum size 0.25\,$\rm \mu m$, and a power-law slope of $-3.5$. The temperature of the dust is taken from the MHD simulation, and a perfect alignment of the silicate dust grains is assumed. The ray-tracing is done with parallel pencil beams over an area of 560\,pc by 560\,pc, with the observer position outside of the simulated galaxy and offset $-8$\,kpc in the disk plane to mimic a (doubled) line of sight at $l$\,$=$\,90$^{\circ}$ through the Milky Way. The selected area gives an angular size of 4 arc degrees by 4 arc degrees at the distance of 8\,kpc, and the number of pixels is selected by the pixel size of PLANCK maps (1.72\arcmin, smoothed afterwards with a 5\arcmin\ FWHM Gaussian beam) and PRIMAger, respectively, producing maps of the four Stokes vector components at 235\,$\rm \mu m$ wavelength.

We highlight the additional information obtained with PRIMAger's higher angular resolution by filtering the large scales from the Stokes parameters maps. We computed the second-order structure function of the polarization angles \cite{Houde+2009} derived from the PRIMAger Stokes $Q$ and $U$ and identified the spatial scales dominated by the mean orientation, which appear around 20\arcmin. Then, we convolved the PRIMAger Stokes $Q$ and $U$ maps with a Gaussian beam and subtracted the result from the initial maps, effectively high-pass filtering PRIMAger maps.
This operation is possible thanks to the broader dynamic range in the PRIMAger observations. A similar process in the Planck data is unavoidably limited by the number of independent beams that can be used for the mean-field orientation. 

\begin{figure*}[ht]
\includegraphics[width=\textwidth,angle=0,origin=c]{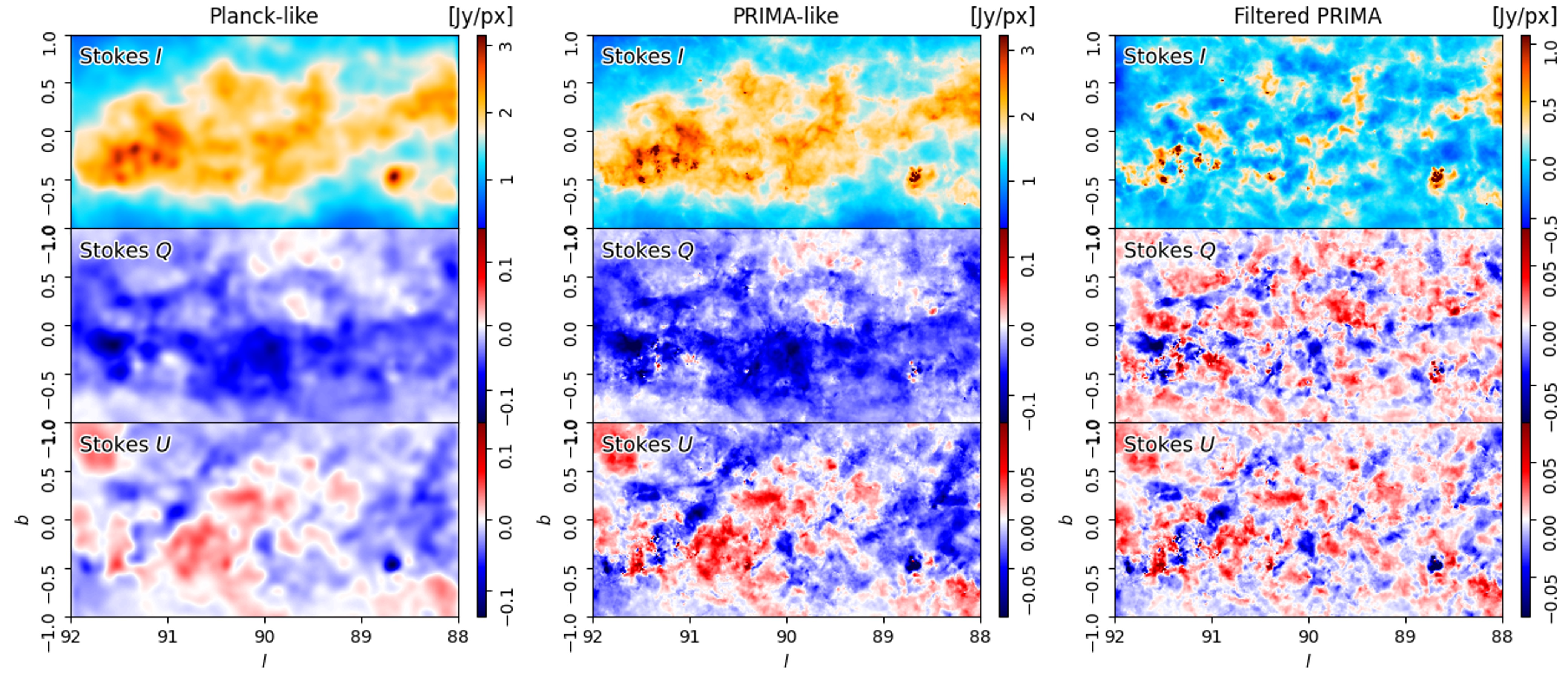}
\caption{Segment of the synthetic observations of the Galactic plane obtained from the numerical simulations of a Milky-Way-like galaxy (see text) processed with {\tt Polaris}. Stokes I, Q and U parameters are shown top to bottom for maps at Planck resolution (left), PRIMAger resolution (center), and filtered PRIMAger (right). }
\label{fig:SynthObs}
\end{figure*}

\begin{figure*}[hb]
\includegraphics[width=1\textwidth,angle=0]{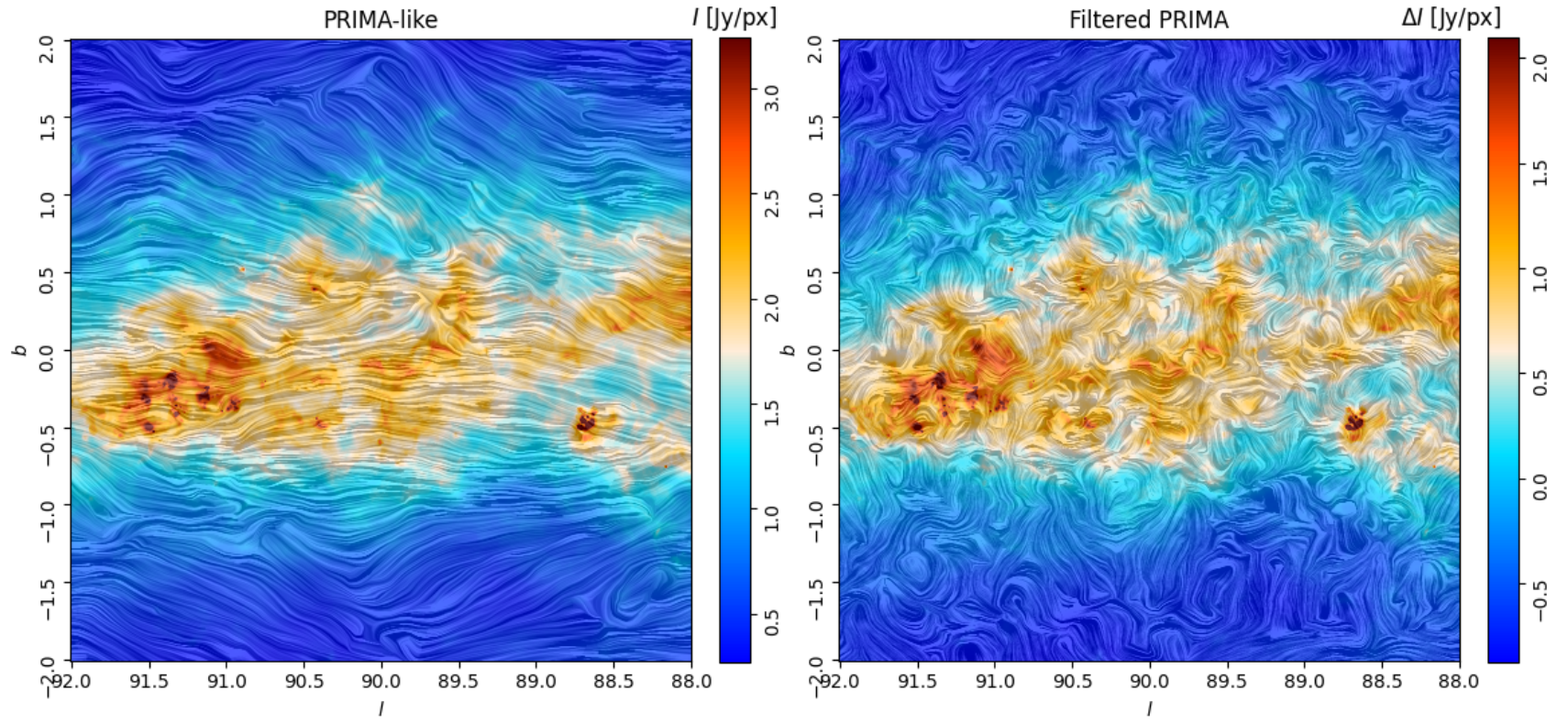}
\caption{LIC representation of the magnetic field pattern in PRIMAger simulated maps (left), and after filtering the large-scale extended component (right).}
\label{fig:prima_lic}
\end{figure*}

Figure \ref{fig:SynthObs} shows the result of this first exercise. In the left column we produce synthetic maps of I, Q and U Stokes parameters (top to bottom) at the Planck $\sim$5\arcmin\ resolution for comparison. The dominant feature is the diffuse positive emission in Stokes Q, that explains the dominant horizontal magnetic field orientation observed. The same pattern is seen at the PRIMAger resolution of $\sim$20\arcsec\ (central column). Thanks to the much richer emission at scales not accessible to Planck, as explained above we are now in the condition to effectively filter this extended component. The result is shown in the third column, and proves that the dominant horizontal pattern is due to the extended large-scale ISM emission along the line of sight. Stokes Q and U parameters in the filtered maps reveal a rich and complex structure that effectively traces the polarization emission of the denser structures in the maps, demonstrating that polarization observations with PRIMAger will be a fundamental tool effective in finally mapping the structure of magnetic fields in and around large-scale filamentary structures throughout our Galaxy in a statistically significant fashion. 
Fig. \ref{fig:prima_lic} shows a Line Integral Convolution (LIC) representation of the direction of the magnetic field obtained from the synthetic data shown in the right panel of Fig. \ref{fig:SynthObs}. The two panels show the results at the PRIMA resolution before (left panel) and after (right panel) filtering of the large spatial scale signal, revealing the complexity of the magnetic field pattern that can be recovered with PRIMA.

It is our plan in the near future to develop this simulations into a full end-to-end pipeline with which we will test lines of sight of variable levels of confusion, different large-scale filtering techniques, and different wavelengths. Indeed, in addition to unprecedented spatial resolution and mapping speed for polarized emission, PRIMAger offers the unique opportunity to simultaneously acquire polarization in 4 different bands between 96 and 235\um. This capability offers additional tools to investigate the relationship between magnetic fields and the density distribution of the ISM along the line of sight. Planck has indeed observed spatial changes in the SED of the polarized component of dust emission that can be related to changes in the coupling between magnetic fields and dust density \cite{Montier-Planck2017}. 
We have demonstrated that the spatial resolution accessible to PRIMAger will enable us to effectively study magnetic field structure in the filamentary cloud where dense 1 pc clumps form, something that was not possible with Planck. This power of PRIMAger can be deployed over the entire Milky Way to provide unprecedented statistical significance.

\section{The PRIMAger polarization Galactic Plane Survey}

\subsection{A large-scale multi-parametric study of the role of magnetic fields in Galactic star formation}

We will use PRIMAger dust polarization maps of the Galactic Plane to measure the magnetization properties of entire molecular filaments/clumps structures, and correlate the magnetization with local star formation rates,  efficiencies and cloud/clump fragmentation degree derived from far-IR continuum and spectroscopic large-scale surveys at comparable resolutions such as Hi-GAL \cite{Molinari+2016a,Elia+2021}, SEDIGISM \cite{DuarteCabral+2021} and FUGIN \cite{Umemoto+2017}, enabling, for the first time the development of a quantitative statistical picture of the role of the magnetic fields in modulating cloud and star formation efficiencies in our Galaxy. 

 We will address the great diversity of density, star formation activity levels, turbulence properties, evolutionary stages and environmental conditions that the star-forming filamentary structures span in our Galaxy. Our target fields will overlap with the sample of 
 1,000 dense clumps surveyed in the ALMAGAL large program \cite{Molinari+2025}, which constitutes the most complete survey of the clump-to-core fragmentation process in our Galaxy. Our polarization observations will enable studies of how magnetic fields affect the local and global infall of the core-hosting gas clumps, and whether magnetic fields inhibit clump fragmentation and remove angular momentum.



\begin{figure*}[ht]
\centering
\includegraphics[width=1.0\textwidth]{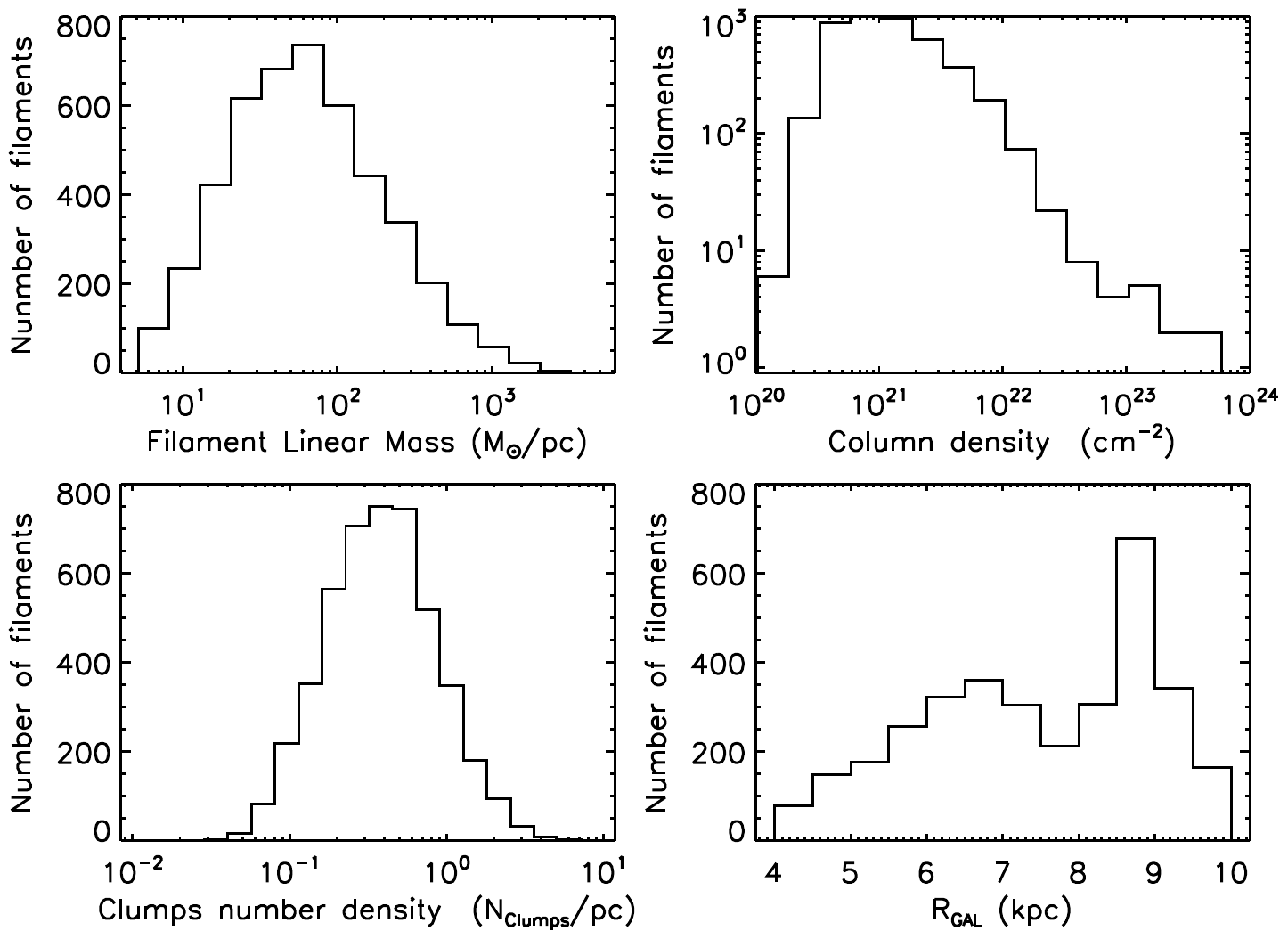} 

\caption {\label{fil_params} Distribution of four key properties of dense filamentary cloud structures within a 640 sq. deg. region of the Galactic Plane with 20\adeg$\leq l \leq$ 340\adeg and $\lvert b\rvert \leq$ 1\adeg, from the {\em Herschel} infrared Galactic Plane Survey catalog \cite{Schisano+2020}: Linear mass, which determines whether filaments should be gravitationally unstable; average background-subtracted filament column density; number density of dense clumps along filamentary structures, that measures the efficiency of filaments fragmentation during their evolution; distance from the Galactic center to the filament ($R_{GAL}$), which probes different Galactic environments.} 
   \hrule
\end{figure*}

The PRIMAger Galactic plane survey will be able to survey several thousands of dense filamentary clouds within 8,000\,pc distance, for which it will make maps at $<$\,0.4\,pc resolution (assuming the more distant structures at 8 kpc from the Sun). Current far-IR continuum (Hi-GAL) and CO spectroscopic (SEDIGISM) surveys show fields with thousands of dense filamentary clouds that have little or no line-of-sight confusion over Galactic longitude ranges, e.g. between 355$^{\circ}$ and 220$^{\circ}$. With such a large survey we will fully sample the 4-dimensional parameter space of important filament properties shown in Figure \ref{fil_params}. 
We aim to provide a statistically significant Galaxy-scale quantification of the role of magnetic fields in the formation of large-scale filaments and the subsequent fragmentation into dense clumps. The Herschel Hi-GAL survey provides a complete census of filamentary structures of more than 30,000 candidates automatically extracted over the entire Galactic Plane \cite{Schisano+2020} located from the Central Molecular Zone out to the external regions of the Milky Way. As we want to concentrate on long consistent structures, a selection of filaments i) longer than the Taurus B211/B213 filament ($\sim$3 pc), ii) longer than 5 armin to provide sufficiently sampled structures and iii) closer than 8 kpc to avoid excessive beam dilution of the structures, provides a sample of $\sim$5000 filaments. 

Fig. \ref{fil_params} shows the distribution of 4 parameters characterising fundamental properties the above selected Hi-GAL sample of filamentary structures. The linear mass defines the regime of stability accessible to this survey. The average column density may be an evolutionary indicator as well as a tracer of the efficiency of the filament formation mechanism. Hi-GAL also fully characterised the dense clumps that fragment along filaments \cite{Elia+2021}; the bottom-left panel of the figure report the number of dense clumps per unit length on filaments, and it is another crucial parameter to correlate with magnetic field properties. The Galactocentric distance is a proxy for the diversity of the environments in which filaments form in the Galactic ecosystem. Finally (not shown in the figure), the distribution of L/M of the dense clumps, defined a the ratio between the bolometric luminosity of a dense clump and its mass found along the filaments, that may trace filament evolution as well as the efficiency of the fragmentation mechanisms. 

Integrated masses span over four orders of magnitude from 10 to 10$^5$\msun.
The average filament column density ranges over four orders of magnitude from 10$^{19}$ to 10$^{23}$\cmtwo, which includes the transition from a magnetically supported cloud (with a constant B over N$_H$) to a supercritical regime (over $\sim$10$^{22}$\cmtwo) where gravity is dominant and B grows with N$_H$ \cite{Crutcher+2010}. Finally, the filaments sample extends from the Bar region to the Outer Arm, crossing the Norma, Scutum-Crux, Sagittarius-Carina, and Perseus Arms (and, of course, the respective interarm regions). 

The variance offered by the sample offers a continuum of situations, from structures with low column densities and virtually no fragmentation into dense clumps (or clumps in a relatively early stage with L/M$\leq$1), where we will be able to map magnetic fields during the filament assembly process, to very dense, supercritical and fragmented structures where we will map the properties of magnetic fields against the rates and efficiencies of the ongoing star formation process.


PRIMAger's multiband mapping capabilities will provide additional tools to disentangle the coupling of magnetic fields and dust properties to polarized emission. For the more diffuse phase, for column densities below few 10$^{21}$ \cmtwo, the measurement of the spatial decorrelation of polarised emission at different wavelengths will provide important clues in this respect \cite{Montier-Planck2017}. For the denser phase of the ISM along filaments, the emission starts becoming optically thick for decreasing wavelength, so we can measure polarization at different depths along the line of sight towards the dense structures, therefore performing a limited tomography of the magnetic field structure. The instrument features four different detector arrays (Ciesla et al., this Volume) closely packed so that their respective footprints are all contained within an $\sim$11\arcmin $\times$11\arcmin\ area. With our proposed large-scale mapping (see below) this arrangement will effectively deliver simultaneous 4-band polarization maps in a single mapping observations.

\subsection{From polarization to strength and morphology of magnetic fields}

 To measure the magnetic properties of the clouds for this unprecedented volume of multi-scale polarimetry observations, not available from any other instrument, we will deploy an integrated set of state-of-the-art statistical tools.
 
 We will first constrain the dust grain alignment efficiency and the inclination angle of the magnetic field by exploring the dependence of the polarization fraction ($p$) on the local dispersion ($S$) in the field orientation angle, and the gas column density ($N$) using the method called {\em $pNS$ relations} \cite{Fissel+2016}

We will 
then estimate the plane-of-sky magnetic field strength component using the Polarization Dispersion Analysis technique \cite{Houde+2009}. PDA, a refinement of the  Davis-Chandrasekhar-Fermi method \cite{Davis1951,CF1953}, uses large numbers of independent polarization vectors to study dispersion of the magnetic field direction vs.~scale. Combined with the velocity information obtained from existing and forthcoming molecular line observations at similar spatial resolutions, and field inclination angle estimates from the pNS relations, the PDA will provide estimates of the magnetic field strength, turbulence power spectrum, 
and energy balance between magnetic pressure, turbulence, and self-gravity.

Finally, the PDA-derived magnetic field properties will be tested by the analysis of the coupling between the density and magnetic fields with the Histogram of Relative Orientations technique \cite{Soler+2013} and directional analysis metrics like the projected Rayleigh statistic \cite{Jow+2018,Soler2019}, which by quantifying the relative orientation between cloud density structures and the magnetic field constrains the magnetic to turbulence energy density ratio. HRO magnetization estimates (which were at the core of the {\sl Planck} polarization analysis discussed above) are independent of those made using the PDA technique, thus providing a complementary constraint on the magnetic energy density. 

\subsection{Time estimates}

The filamentary structures detected in the Herschel Hi-GAL maps \cite{Schisano+2020} have a typical brightness from several tens to thousands of MJy/sr at 250$\mu$m, that basically translates into a corresponding range of mean $N_H$ columns densities that is shown in the top-right panel of Fig.\ref{fil_params}. By adopting 100 MJy/sr as a working figure for detection of continuum surface brightness we make sure that we will be able to map structures characterized by $A_V\leq 1$, and hence allow us to investigate the formation of high-density filaments from low-density molecular gas. Since we would like to detect a 100 MJy/sr minimum brightness at least 3\% of the continuum as polarized signal to trace depolarization across the filaments (as observed, e.g., for nearby Gould Belt filaments) at a 5$\sigma$ significance, our required sensitivity is 1 MJY/sr. With the  PRIMAger instrument sensitivity (see L. Ciesla in this Volume) a brightness of 1.25 MJy/sr can be reached over a mapped area of 1 sq. deg. at a 5$\sigma$ level in 10 hours in Large Mapping mode including all estimated overheads. Note that this estimate also accounts for the two orthogonal scan passes needed to calibrate and remove the 1/f noise of the detectors.

Our required sensitivity of 1 MJy/sr \@1$\sigma$ will be reached $\sim$1.7 hours for a one sq. deg. map. Mapping the entire Galactic Plane in a $\lvert$b$\rvert\leq 1$\adeg strip, a 720 sq. deg. total area, requires $\sim$1200 hours total observing time.

\section{Conclusions}

PRIMAger has the potential to be a game-changer to reconstruct the structure of the magnetic fields towards the Milky Way Galactic Plane that, at Planck resolutions, is dominated by the emission from diffuse ISM that washes out the contribution from dense material. We demonstrate the feasibility of an extensive Galactic Plane survey with PRIMAger, recovering polarization and hence magnetic field from several thousands of filamentary structures in different stages of evolution, to finally address in a statistically significant fashion the question of the role of the magnetic field in the formation, assembly, and evolution of dense star-forming clouds.


\subsection* {Acknowledgments}
The authors gratefully acknowledge financial support from the European Research Council via the ERC Synergy Grant ``ECOGAL'' (project ID 855130).

\subsection*{Code and data availability statement}
This paper does not make use of any instrument data. The processing done to produce figures 1 and 2 uses post-processing done with the publicly available POLARIS code \cite{Reissl+2016} running on numerical simulations that are yet unpublished.

\subsection*{Disclosure statement}
The authors declare there are no financial interests, commercial affiliations, or other potential conflicts of interest that have influenced the objectivity of this research or the writing of this paper.

\bibliographystyle{spiejour}   

\bibliography{sergio_bib_2}   

\begin{thebibliography}{10}

\bibitem{Crutcher+2010}
R.~M. {Crutcher}, B.~D. {Wandelt}, C.~{Heiles}, E.~{Falgarone}, and T.~H. {Troland}, ``Magnetic fields in interstellar clouds from zeeman observations: Inference of total field strengths by bayesian analysis,'' {\em \apj} {\bf 725}, 466  (2010).

\bibitem{Schisano+2014}
E.~{Schisano}, K.~L.~J. {Rygl}, S.~{Molinari}, G.~{Busqu\'et}, D.~{Elia}, M.~{Pestalozzi}, D.~{Polychroni}, N.~{Billot}, S.~J. {Carey}, R.~{Paladini}, and {et al.}, ``The identification of filaments on far-infrared and submillimiter images: Morphology, physical conditions and relation with star formation of filamentary structure,'' {\em \apj} {\bf 791}, 27  (2014).

\bibitem{Schisano+2020}
E.~{Schisano}, S.~{Molinari}, D.~{Elia}, M.~{Benedettini}, L.~{Olmi}, and {et al.}, ``The hi-gal catalogue of dusty filamentary structures in the galactic plane,'' {\em \mnras} {\bf 492}, 5420  (2020).

\bibitem{Molinari2010b}
S.~{Molinari}, B.~{Swinyard}, J.~{Bally}, M.~{Barlow}, J.-P. {Bernard}, P.~{Martin}, T.~{Moore}, A.~{Noriega-Crespo}, R.~{Plume}, L.~{Testi}, A.~{Zavagno}, A.~{Abergel}, B.~{Ali}, L.~{Anderson}, P.~{Andr\'e}, J.-P. {Baluteau}, C.~{Battersby}, M.~T. {Beltr\'an}, M.~{Benedettini}, N.~{Billot}, J.~{Blommaert}, S.~{Bontemps}, F.~{Boulanger}, J.~{Brand}, C.~{Brunt}, M.~{Burton}, L.~{Calzoletti}, S.~{Carey}, P.~{Caselli}, R.~{Cesaroni}, J.~{Cernicharo}, S.~{Chakrabarti}, A.~{Chrysostomou}, M.~{Cohen}, M.~{Compiegne}, P.~{de Bernardis}, G.~{de Gasperis}, A.~M. {di Giorgio}, D.~{Elia}, F.~{Faustini}, N.~{Flagey}, Y.~{Fukui}, G.~A. {Fuller}, K.~{Ganga}, P.~{Garcia-Lario}, J.~{Glenn}, P.~F. {Goldsmith}, M.~{Griffin}, M.~{Hoare}, M.~{Huang}, D.~{Ikhenaode}, C.~{Joblin}, G.~{Joncas}, M.~{Juvela}, J.~M. {Kirk}, G.~{Lagache}, J.~Z. {Li}, T.~L. {Lim}, S.~D. {Lord}, M.~{Marengo}, D.~J. {Marshall}, S.~{Masi}, F.~{Massi}, M.~{Matsuura}, V.~{Minier}, M.-A. {Miville-Desch{\^e}nes}, L.~A. {Montier}, L.~{Morgan}, F.~{Motte}, J.~C.
  {Mottram}, T.~G. {M{\"u}ller}, P.~{Natoli}, J.~{Neves}, L.~{Olmi}, R.~{Paladini}, D.~{Paradis}, H.~{Parsons}, N.~{Peretto}, M.~{Pestalozzi}, S.~{Pezzuto}, F.~{Piacentini}, L.~{Piazzo}, D.~{Polychroni}, M.~{Pomar{\`e}s}, C.~C. {Popescu}, W.~T. {Reach}, I.~{Ristorcelli}, J.-F. {Robitaille}, T.~{Robitaille}, J.~A. {Rod{\'o}n}, A.~{Roy}, P.~{Royer}, D.~{Russeil}, P.~{Saraceno}, M.~{Sauvage}, P.~{Schilke}, E.~{Schisano}, N.~{Schneider}, F.~{Schuller}, B.~{Schulz}, B.~{Sibthorpe}, H.~A. {Smith}, M.~D. {Smith}, L.~{Spinoglio}, D.~{Stamatellos}, F.~{Strafella}, G.~S. {Stringfellow}, E.~{Sturm}, R.~{Taylor}, M.~A. {Thompson}, A.~{Traficante}, R.~J. {Tuffs}, G.~{Umana}, L.~{Valenziano}, R.~{Vavrek}, M.~{Veneziani}, S.~{Viti}, C.~{Waelkens}, D.~{Ward-Thompson}, G.~{White}, L.~A. {Wilcock}, F.~{Wyrowski}, H.~W. {Yorke}, and Q.~{Zhang}, ``{Clouds, filaments, and protostars: The Herschel Hi-GAL Milky Way},'' {\em A\&A} {\bf 518}, L100  (2010).

\bibitem{Soler-Planck2016}
{Planck Collaboration}, ``Planck intermediate results. xxxv. probing the role of the magnetic field in the formation of structure in molecular clouds,'' {\em \aap} {\bf 586}, 138  (2016).

\bibitem{Soler+2013}
J.~D. {Soler}, P.~{Hennebelle}, P.~G. {Martin}, M.-A. {Miville-Desch{\^e}nes}, C.~B. {Netterfield}, and L.~M. {Fissel}, ``An imprint of molecular cloud magnetization in the morphology of the dust polarized emission,'' {\em \apj} {\bf 774}, 128  (2013).

\bibitem{Friberg+2016}
P.~{Friberg}, P.~{Bastien}, D.~{Berry}, G.~{Savini}, S.~F. {Graves}, and K.~{Pattle}, ``Pol-2: a polarimeter for the james-clerk-maxwell telescope,'' {\em Proc. SPIE} {\bf 9914}  (2016).

\bibitem{Harper+2018}
D.~A. {Harper}, M.~C. {Runyan}, C.~D. {Dowell}, C.~J. {Wirth}, M.~{Amato}, and {et al.}, ``Hawc+, the far-infrared camera and polarimeter for sofia,'' {\em J. Astr. Instr.} {\bf 7}  (2018).

\bibitem{Pillai+2020}
T.~{Pillai}, D.~P. {Clemens}, S.~{Reissl}, P.~C. {Myers}, J.~{Kauffmann}, and {et al.}, ``Magnetized filamentary gas flows feeding the young embedded cluster in serpens south,'' {\em Nature Astronomy} {\bf 4}, 1195  (2020).

\bibitem{Liu+2018}
T.~{Liu}, P.~S. {Li}, M.~{Juvela}, K.~T. {Kim}, N.~J. {Evans}, II, and {et al.}, ``A holistic perspective on the dynamics of g035.39-00.33: The interplay between gas and magnetic fields,'' {\em \apj} {\bf 859}, 151  (2018).

\bibitem{Bernard+2015}
{Planck Collaboration}, ``Planck intermediate results. xix. an overview of the polarized thermal emission from galactic dust,'' {\em \aap} {\bf 576}, 104  (2015).

\bibitem{Tress+2024}
R.~G. {Tress}, M.~C. {Sormani}, P.~{Girichidis}, S.~C.~O. {Glover}, R.~S. {Klessen}, and {et al.}, ``Magnetic field morphology and evolution in the central molecular zone and its effect on gas dynamics,'' {\em \aap} {\bf 691}, 303  (2024).

\bibitem{Reissl+2016}
S.~{Reissl}, S.~{Wolf}, and R.~{Brauer}, ``Radiative transfer with polaris. i. analysis of magnetic fields through synthetic dust continuum polarization measurements,'' {\em \aap} {\bf 593}, 87  (2016).

\bibitem{Houde+2009}
M.~{Houde}, J.~E. {Vaillancourt}, R.~H. {Hildebrand}, S.~{Chitsazzadeh}, and L.~{Kirby}, ``Dispersion of magnetic fields in molecular clouds. ii.,'' {\em \apj} {\bf 706}, 1504  (2009).

\bibitem{Montier-Planck2017}
{Planck Collaboration}, ``Planck intermediate results l. evidence of spatial variation of the polarized thermal dust spectral energy distribution and implications for cmb b-mode analysis,'' {\em \aap} {\bf 599}, 51  (2017).

\bibitem{Molinari+2016a}
S.~{Molinari}, E.~{Schisano}, D.~{Elia}, M.~{Pestalozzi}, A.~{Traficante}, S.~{Pezzuto}, B.~{Swinyard}, A.~{Noriega-Crespo}, and {et al.}, ``Hi-gal, the herschel infrared galactic plane survey: photometric maps and compact source catalogues. first data release for inner milky way: +68$^{\circ} \geq l \geq -$70$^{\circ}$,'' {\em A\&A} {\bf 591}, 149  (2016).

\bibitem{Elia+2021}
D.~{Elia}, M.~{Merello}, S.~{Molinari}, E.~{Schisano}, A.~{Zavagno}, and {et al.}, ``The hi-gal compact source catalogue - ii. the 360$\,^{\circ}$ catalogue of clump physical properties,'' {\em \mnras} {\bf 504}, 2742  (2021).

\bibitem{DuarteCabral+2021}
A.~{Duarte-Cabral}, D.~{Colombo}, J.~S. {Urquhart}, A.~{Ginsburg}, D.~{Russeil}, and {et. al.}, ``The sedigism survey: molecular clouds in the inner galaxy,'' {\em \mnras} {\bf 500}, 3027  (2021).

\bibitem{Umemoto+2017}
T.~{Umemoto}, T.~{Minamidami}, N.~{Kuno}, S.~{Fujita}, and {et al.}, ``Forest unbiased galactic plane imaging survey with the nobeyama 45 m telescope (fugin). i. project overview and initial results,'' {\em \pasp} {\bf 69}, 78  (2017).

\bibitem{Molinari+2025}
S.~{Molinari}, P.~{Schilke}, C.~{Battersby}, P.~T.~P. {Ho}, and {et al.}, ``Almagal i. the alma evolutionary study of high mass protocluster formation in the galaxy: Presentation of the survey and early results.,'' {\em \aap} {\bf 696}, A149  (2025).

\bibitem{Fissel+2016}
L.~M. {Fissel}, P.~A.~R. {Ade}, F.~E. {Angil\'e}, P.~{Ashton}, S.~J. {Benton}, and {et al.}, ``Balloon-borne submillimeter polarimetry of the vela c molecular cloud: Systematic dependence of polarization fraction on column density and local polarization-angle dispersion,'' {\em \apj} {\bf 824}, 134  (2016).

\bibitem{Davis1951}
L.~{Davis}, ``The strength of interstellar magnetic fields,'' {\em Physical Review} {\bf 81}, 890  (1951).

\bibitem{CF1953}
S.~{Chandrasekhar} and E.~{Fermi}, ``Magnetic fields in spiral arms,'' {\em \apj} {\bf 118}, 113  (1953).

\bibitem{Jow+2018}
D.~L. {Jow}, R.~{Hill}, D.~{Scott}, J.~D. {Soler}, P.~G. {Martin}, M.~J. {Devlin}, L.~M. {Fissel}, and F.~{Poidevin}, ``An application of an optimal statistic for characterizing relative orientations,'' {\em \mnras} {\bf 474}, 1018  (2018).

\bibitem{Soler2019}
J.~D. {Soler}, ``Using herschel and planck observations to delineate the role of magnetic fields in molecular cloud structure,'' {\em \aap} {\bf 629}, 96  (2019).

\end{thebibliography}

\listoffigures
\listoftables

\end{spacing}
\end{document}